# Virtual-organism toy-model as a tool to develop bioinformatics approaches of Systems Biology for medical-target discovery

Albert Pujol *1; Raquel Valls*1, Vesna Radovanovic*1; Emre Guney*2; Javier Garcia-Garcia*2; Victor Codony Domenech*1, Laura Corredor Gonzàlez*1; J .M. Mas*1; Baldo Oliva*2+

*1 Anaxomics Biotech. c/Balmes 89 4º 2ª, 08008 Barcelona, Spain.

*2 Structural Bioinformatics Group (GRIB–IMIM). Departament de Ciències Experimentals i de la Salut. Universitat Pompeu Fabra. C/Doctor Aiguader, 83, 08003 Barcelona, Catalonia, Spain

+ corresponding author: Baldo Oliva

**ABSTRACT**

Systems Biology has emerged in the last years as a new holistic approach based on the global understanding of cells instead of only being focused on their individual parts (genes or proteins), to better understand the complexity of human cells.

Since the Systems Biology still does not provide the most accurate answers to our questions due to the complexity of cells and the limited quality of available information to perform a good gene/protein map analysis, we have created simpler models to ensure easier analysis of the map that represents the human cell. Therefore, a virtual organism has been designed according to the main physiological rules for humans in order to replicate the human organism and its vital functions. This toy model was constructed by defining the topology of its genes/proteins and the biological functions associated to it.

There are several examples of these toy models that emulate natural processes to perform analysis of the virtual life in order to design the best strategy to understand real life. The strategy applied in this study combines topological and functional analysis integrating the knowledge about the relative position of a node among the others in the map with the conclusions generated by mathematical models that reproduce functional data of the virtual organism. Our results demonstrate that the combination of both strategies allows better understanding of our virtual organism even with the lower input of information needed and therefore it can be a potential tool to better understand the real life.

**KEY WORDS:**
- Systems Biology
- Functional Analysis
- Topological Analysis
- Algorithm
- Protein network

## 1. INTRODUCTION

Molecular Biology of the last century was dedicated to understand the mechanisms of action and the regulation of all genes or proteins in order to develop new drugs for specific situations. The main effort was made to learn how a protein or a gene works in the cell.

This effort drove us toward a deeper understanding of some specific mechanisms of action of proteins such as receptors (e.g.: coagulation cascade [1]; G protein-coupled [2]) or even some pathways like KEGG [3]. However, even with genome project been concluded [1, 4], the available information is absolutely insufficient to understand the complexity of a cell. Given this fact a new approach has appeared in the last years based on the global understanding of cells, instead of treating the relevance of their individual parts that are genes and proteins. In this new approach all the available knowledge is considered as a global system where the specific information (proteins and/or genes) are less important than the global relationship between them. This new approach is called Systems Biology (SB) [5] and it aims to represent a cell as a map of interconnected genes or proteins. Each item in this map (gene or protein) is usually represented by a circle and it is considered to be a node. The relation between nodes is represented by a line linking them. There are several types of maps depending on the type of relation between nodes; the interactome is the map that establishes physical relations between proteins [6], metabolome is the map that relates proteins by its functional behavior, and the regulatory map, which establishes relation between genes and/or proteins depending on their regulation [7].

So, SB has transformed proteins and genes into a map of nodes linked by a specific relation between them, allowing the transformation of the cell into a graph. How can we use this representation of the nature in our work? There are two main types of strategies: a) those based on the topological information of the map (Topological strategies), and b) those that include the map but also any other functional information associated to a cell (Functional strategies). Topological methods are based on the analysis of the relative position of a node among the others. By simple intuition we could think that closer nodes are more strongly related than the farther ones, but this is not clear. For instance, the protein Gamma-aminobutyric acid receptor subunit alpha-1 (with P14867 Uniprot code) is an effector of Arthralgia adverse event (AE) [8], however, the drug diazepam [with DB00829 DrugBank code] has this protein as a target and Arthralgia is not reported in the leaflet of this drug [9].

Currenttopological strategies include the analysis of fluxes over the map [10, 11] to detect bottlenecks and hubs, determining these proteins as strategic positions in the map.

Using this approach it is possible to explain certain mechanisms of action of drugs and to identify key points over the map that eventually become key proteins or genes in cells [12-14]. Functional strategies include topological information but they also take into account functional data of the cell. An example of this functional data is the effect of drugs. In this case drug targets are considered to be inputs to the map (being +1 the activation and -1 the inhibition), and the therapeutic indication of a drug and its adverse events are considered as outputs[15]. Inputs and Outputs allow transforming a static map into a dynamic system with apparently more possibilities to give answers closer to the observations of the nature. Topological and Functional approaches help us clarify the mechanisms of action in some cases [16, 17], and for drug repositioning [18, 19] in others. Unfortunately, these methods do not have enough precision and they are not useful to understand all the drugs yet.

Complexity of cells, due to the high number of nodes, and the quality of available data caused by a large number of false relationships between nodes, are the main impediment to analyze the graph provided by SB. Thus, it is necessary to create simpler models to determine the way of analyzing the graph that represents the cell. In this sense, we present a virtual model of a worm with basic vital functions of living, growing up, eating, defecating, breathing and regulating its temperature. The same aspects that we have to understand about human cells can be applied to this virtual worm because it has been designed according to the main physiological human principles. Is it possible to understand how the virtual worm lives? Is it possible to identify virtual worm's key proteins for life? Could we design a drug treatment to cure any pathology associated with any biological activity? To answer these questions both types of strategies have been applied: the topological and the functional.

## 2. METHODS

### *2.1. Creating a virtual life*

To reach the objectives planned for this work it was necessary to define a model of a cell and then to define the adequate mathematical strategy to understand this model. The model of a cell created for this project represents a unicellular virtual worm which satisfies the basic human physiological principles. This virtual worm was designed with basic functions associated to its life: growing up, eating, defecating, breathing, and regulating its temperature. A model of a cell in SB is often represented by a set of genes or proteins linked by lines that define the relation between these genes or proteins. In mathematical terms the gene/protein map is described by a graph containing nodes (genes/proteins) usually represented as circles and the lines linking them. For human cells some genes/proteins belong to known pathways that are associated to biological functions. The virtual model has been designed including pathways as well. So, similarly to human cells, each one of these basic functions associated to the virtual worm has been created as a pathway that contains a set of genes/proteins. The structure and the functionality of the pathways have been created by following similar pathways and known biological functions in humans. Figure 1 describes the four pathways existing in the virtual worm. The graph represents the virtual worm genes/proteins (nodes) and the relations existing between each other (links). Specific proteins of the map are represented as nodes whose names have been assigned by combining the letter P and a number. Arrows are links among proteins and black arrows represent the connections among different pathways. Filled nodes are effector proteins, which are proteins/genes associated to the pathway activity but their function does not have to be necessarily relevant for the pathway. Each pathway is represented by a different color: green color represents the pathway associated with the respiration function, red represents pathway for the feed, yellow represents the pathway associated to the temperature regulation and blue is for the excretion function. The structure of each pathway and the genes or proteins contained in each pathway of the worm have been replicated from the other known pathways in mammalian cells, being cyclic, lineal or in a form of cascades of genes/proteins [3].

For instance, respiration is represented by a linear pathway and excretion is represented by a cycle (Figure 1). Following the patterns of KEGG, some key genes/proteins have been associated with more than one pathway of the virtual worm and therefore they control the cross-activity between pathways of the virtual worm. The same process can be observed in relations between the pathways in human cells. Some other genes/proteins of the map of the virtual worm have the capability to capture influences from the environment: cold, oxygen, food. Having in mind the limitation that exists in humans, only specific genes/proteins of a pathway of the virtual worm have the capability to be measured. Therefore, measures of these proteins/genes are associated to the pathway activity but their function does not have to be necessarily relevant for the pathway. Usually these nodes are at the end of each pathway and they are called **effectors.** Namely, the effector proteins of the gene/protein map of the virtual worm are P8, P14, P24 and P30 and have been represented as nodes filled with colors in Figure 1. Furthermore, the designers of the virtual worm have selected a group of key proteins that are indispensable for the life of the virtual worm under the known premises of life. These proteins are: P3, P5, P11, P23 and P28 (Figure 1) and are considered to be **seed proteins**. The seed proteins are those with a relevant role in any of the four essential pathways of the virtual worm's life. They have been selected because all of them connect to different pathways or are highly interconnected. These proteins were selected by the virtual worm designer before starting any analysis and they are crucial to carry out the Topological analysis since the map will be built around these seed proteins.

While Figure 1 summarizes the genes/proteins of the virtual worm from a topologic point of view, the Figure 2a includes a table with the functional principles of life of the virtual worm. The table shows a portion of the entire truth-table. Sixty one conditions, compatible with the life of the virtual worm, have been described. An external signal (green part of the table) or a combination of signals produce a response (white part of the table) in the virtual worm activating (green color) or inhibiting (red color) a metabolic pathway or a protein. White color represents unknown response. So, the biological functions associated with the virtual worm have been described in the Figure 2a as a Truth-Table including information about the respiration, temperature regulation, feed and excretion. Taking into account the same information that is known for humans, the Truth-Table contains definitions about the 4 pathways of the virtual worm and how these pathways interact with the environment. An example of the type of information contained in Figure 2a is the following: in order to grow up it is necessary to eat; at the same time, eating enables temperature regulation and it is necessary for breathing which requires oxygen. Figure 2a contains 61 rules similar to these ones that explain how the virtual worm lives, for instance, column 5 indicates that presence of food activates the possibility of feeding and wasting. The information that is contained in Figure 2a has 2 levels of deepness: a) specific information about proteins of the virtual worm, and b) the abstract information that includes no specific data about proteins. This effect of the two types of deepness is equivalent to the information that we have about humans where some proteins are well known while other information is only associated with pathways or biological functions, but not with specific proteins. So, the abstract information thoroughly describes the biological activity of the virtual worm, for instance, the result of eating is growing up. The abstract information cannot be represented in Figure 1, for example, it is impossible to represent the process of growing up in the map. This fact has forced us to introduce 4 new nodes (Figure 2b) that are representing the abstraction of the function of each pathway in the topological map. Figure 2b shows the virtual worm genes/protein map representing the pathways associated with respiration (in green), with feeding (in red), with temperature regulation (in yellow) and with excretion (in blue). Smaller filled nodes are effector proteins and the 4 biggest filled nodes correspond to the abstraction points for the functional analysis.

Therefore, the definition of the virtual worm is contained in Figure 1 (Topological) and Figure 2 (Functional).

### *2.2. How to understand the life of the virtual organism?*

There are two different strategies to analyze what the life of this virtual worm will be like. The first strategy, called Topological analysis, is based on the relative position of genes/proteins of the virtual worm and the links between them. The second type of analysis, known as Functional analysis, is based on the information related to the interaction of the virtual worm with its environment. To perform both analysis, Topological and Functional, certain information is required. Figure 3 shows the available and hidden information and the results and validation of the topological and functional analyses. The seed proteins* mentioned in figure 3 are proteins with a relevant role in any of the four essential pathways of the virtual worm's life [15]. The effector proteins** mentioned in figure 3 are proteins with important role in the map and whose activity is directly related to the phenotype [20].

In the case of the Topological analysis there are two requirements: a) the topology of genes/proteins of the virtual worm that is presented in the Figure 1, and b) seed proteins associated with the map (Figure 3). These proteins are usually the ones with a relevant role in any of the four basic pathways of worm's life. On the other hand, Functional analysis has two requirements (Figure 3): a) the topology of genes/proteins of the virtual worm (Figure 1), and b) the Truth-Table that contains the functional explanation about the virtual worm's life (Figure 2a).

To perform the analysis of both strategies the information required by the first strategy is hidden for the second and vice versa, being especially remarkable that seed proteins were hidden for the analysts of the Functional strategy (Figure 3).

### *2.3. Topological analysis of the genes/proteins of the virtual worm: the virtual worm itself*

Topological analysis studies the structure of genes/proteins of the virtual worm; the virtual worm itself. Therefore, the Topological analysis requires the virtual worm's topological map of genes/proteins which defines its life. This analysis also requires a preliminary list of seed proteins (Figure 3), since the relative position between these and the rest of the proteins determines the importance of each protein in the map allowing the identification of new relevant proteins which correspond to effector proteins. In this case, the typical analysis consists of searching the map for hubs, which are highly connected proteins, and bottlenecks, which are nodes with the highest number of pathways going through them. Therefore, according to their definitions, hubs and bottlenecks control most of the information flow in the network, representing the critical points of the network and subsequently, representing key proteins in the virtual worm's life.

Therefore, the Topological analysis requires the topological map and a preliminary list of seed proteins that are provided by the designers of the virtual worm. Finally, the Topological Analysis allows the identification of novel key proteins in the virtual worm's life which correspond to effector proteins (Figure 3).

### *2.4. Functional analysis of the virtual worm: the virtual worm in its environment*

The Functional analysis studies the relation of the virtual worm with the environment and how the virtual worm adapts its genes/proteins to its life conditions. In other terms, this analysis will consider the information described in the Truth-Table of the Figure 2a but also the topology of genes/proteins in the map. So, the Functional analysis takes into account both the topology of the map (Figure 1) and the functional knowledge about the virtual worm's life (Figure 2a).

The objective of the Functional analysis is to create a mathematical algorithm that complies with all the Truth-Table information with the restriction given by the topology of genes/proteins of the virtual worm. The optimization method to determine this algorithm has been obtained by using artificial intelligence approaches and more specifically by genetic algorithms. To apply this methodology first it was necessary to define the characteristics of the virtual worm's chromosome. The virtual worm's chromosome contains all genes/proteins that are represented in the map of the Figure 1 and the relations of these genes/proteins between them. So, each link represented in Figure 1 has been considered as a gene of the virtual worm's chromosome. The Figure 1 contains 71 links that correspond to the same number of genes considered in the virtual worm's chromosome. To simplify the problem of finding the mathematical algorithm associated with the life of the virtual worm it was considered that all links have the same transactional function but with different biases. The transactional function is a tri-stable function that integrates all inputs received from the linked nodes in one. So, the different behavior of each link is determined by the bias of the function. In other terms, the biases of the functions that are associated with each gene/protein of the virtual worm's chromosome are the only parameters that define the mathematical model related to the virtual worm. Biases are calculated within the range [-1, 1]. The optimization method to determine the value of the biases for the virtual worm's chromosome is a Genetic Algorithm that was implemented in C# and Matlab [21, 22]. These programs were prepared to perform in pseudo-parallel machines in order to increase the optimization of the time of calculation.

Following the basic rules of the Genetic Algorithm the generation zero was created with 1000 randomly created virtual worms. In other words, the values of the biases of each gene of the chromosome of each one of these 1000 worms were assigned randomly. The fitness function evaluates the accuracy of each virtual worm as a distance between the results of the virtual chromosome and the values of the Truth-Table. The ideal solution would be the distance zero between a virtual worm of this generation and the Truth-Table values. If the distance obtained in the fitness function between a virtual worm and the Truth-Table is zero it means that this specific virtual worm fits in exactly with all the information of the Truth-Table. Of course, the generation zero has few possibilities to fit with the Truth-Table given that the biases of the 71 genes of the 1000 virtual worms of this generation were created randomly. So as to find the solution it is necessary to create several generations of virtual worms. The chromosomes of the virtual worms undergo mutations and subsequently the recombination of their genes in order to have descendants .Each new generation of virtual worms is evaluated and the virtual worms with the best fitting values are considered the survivors. The virtual worms with the worst values are excluded (considered to be dead) and they have no descendents. This process was applied during 7000 generations, where around 3.5 millions of virtual worms have been evaluated. The process was stopped in generation 7000 when 1905 virtual worms were alive. The fitness of these virtual worms of the 7000th generation is between 89% and 91% of values of the Truth-Table, being this numbers the distance given by the fitness function. Instead of considering the best virtual worm as a representative of the virtual worm all living virtual worms were considered to be the solution which provides the complete understanding on the worm's life. So, the average of all the conclusions obtained from each living worm will be considered as the final solution [15].

3. RESULTS

To understand how the virtual worm lives is equivalent to discover what the key genes/proteins are and how the virtual worm interacts with its environment. There are two types of relevant genes/proteins: a) those proteins that are important for the life of the worm (seed proteins), and b) those genes/proteins that are closely related to the pathways of the virtual worm (effector proteins).

The objective of the Topological analysis is to identify effector proteins in the virtual worm's chromosome, or in other words, in the gene/protein map. The technique employed to perform the Topological analysis requires as inputs the identification of some seed genes/proteins. These seed genes/proteins have been previously assigned by the virtual worm designer, taking into account that these proteins could also be relevant for the virtual worm's life, being P3, P5, P11, P23 and P28 the selected seed genes/proteins by the virtual worm designers. So, the objective of the Topological analysis is to determine effector gens/proteins that are also relevant for the virtual worm apart from these already known seed proteins. It is obvious that knowing the seed proteins is a limitation factor of the technique employed, however this is quite common available information that any investigator has about the species they are studying. For instance, it is rather common to know the relevant genes/proteins of human pathways following the KEGG definition. Using the Topological analysis the effector proteins obtained were P8, P14, P24, and P30. Figure 4a shows these effector proteins according to the topological analysis as circled nodes, and these coincide with effector proteins previously assigned by the virtual worm designer. The color scale indicates the certainty grade that a given protein is an effector protein. Nodes filled in green are effector proteins with high degree of certainty and those filled in red are proteins with low degree of certainty as effector proteins. Therefore, the effector proteins obtained from the Topological analysis coincide with the effector proteins previously defined by the virtual worm designer.

It is necessary to perform the Functional analysis in order to determine the genes/proteins effectors associated to the pathways (seed proteins). In the Figure 2b there are 4 big nodes filled with color. These nodes are not associated with any known function or gene/protein of the virtual worm but they are artificially introduced abstract nodes which represent the pathway itself. Without introducing these artificial and abstract nodes it is impossible to introduce information from the Truth-Table such as global information about the pathways. This is because it is not possible to assign the functional activity of a pathway to a specific gene/protein. Each of these abstract nodes is linked to all the nodes of their corresponding pathway (Figure 2b). The weight of the links (bias value) between the nodes of the pathway and the abstract node is a measure of the influence of each node in the pathway. In fact, this is the definition of the effector proteins: genes/proteins whose activities are relevant for their corresponding pathways. In humans, for instance, the effector proteins are those proteins that regulate pathways or produce a directly measurable phenotype. Examples of effector genes/proteins in humans are the caspases in the apoptotic pathway or other enzymes as phospholipase C in signal transduction pathways. From the definition of the topological analysis it is clear that these results will depend on the input information which must be reported by the investigator. Figures 4c, d, e and f show the gene/protein map where the effector proteins of each pathway according to the virtual worm designer and topological analysis that match with effector proteins predicted by the functional analysis are represented as a circle. In three of four pathways, functional analysis has reached the same results as the topological analysis and the effectors defined by the virtual worm designer. In the Temperature control pathway, effector protein was ranked in the third position and the effector protein of this pathway according to the virtual worm designer is represented with an arrow. The color scale indicates the certainty grade that a given protein is a seed protein. Nodes filled in green are seed proteins with high degree of certainty and those filled in red are proteins with low degree of certainty as seed proteins. As shown in Figures 4c, d, e and f, the 75% of the effectors found through the Functional analysis match with the effector gene/proteins described in Figure 2b (colored nodes) and are marked in Figures 4c, d, e and f. The effectors described in Figure 1 were selected by the worm designers taking exclusively into account their knowledge about the virtual worm, being this information hidden to the Functional analysts (Figure 3). It is also remarkable that Functional analysis does not require pre-selected proteins as a starting point. In this analysis the conclusion is directly obtained from the mathematical models corresponding to 7000$^{th}$ generation of virtual worms and most specifically the weight of the biases of specific links.

From the mathematical algorithm obtained through the Functional analysis it is also possible to determine what the key genes/proteins could be for the virtual worm (the seed proteins). In this case, the criterion employed takes into account not only the weight of the links but also the proximity of the rest of the links. This analysis has been done by simulating the inputs of perturbation over the map and analyzing how this perturbation is fluxing across the map. The inputs were given randomly in different nodes of the map with random values within the range [-1,+1]. The input signal of each node is obtained as an integration of all inputs-links over the node. The total of the perturbation is calculated in the tri-stable function intrinsic to each node, being this result the output of this node. Following this procedure, genes/proteins P3, P11, and P23 were obtained as relevant genes/proteins for the flux of information over the map (seed proteins). Figure 4b shows these seed proteins as circled nodes according to the functional analysis. These three proteins are included in the group of five proteins selected as seed proteins by the worm designers, being of course hidden the original information for the analyst of the Functional method. The color scale of figure 4b indicates the certainty grade that a given protein is a seed protein. Nodes filled in green are seed proteins with high degree of certainty and those filled in red are proteins with low degree of certainty as seed proteins.

In a deeper analysis of the mathematical algorithm it is possible to determine that P23 is necessary for the oxidation of fatty acids and amino acids (see Figures 4c, d, e and f). The presence of food and oxygen activates the pathways of respiration and feed. Under these conditions, P11 is being produced, being its essential role to synthesize P20, P21 and P22 metabolites of the excretion pathway. Without P3, other nodes like P27

(temperature control pathway), P16 and P17 (feed pathway) would not be activated (see Figures 4c, d, e and f). Therefore, it is estimated that if some of these proteins do not translate their inputs (perturbation) correctly to the map it will lead to a death of the virtual worm, or in the best case, to cause certain deficiencies in its metabolic pathways.

# 4. DISCUSSION

## *4.1. Conclusions*

This project recreates the life of a virtual worm that was defined exclusively by the relation between its genes/proteins (Figure 1) and its way of living (Figure 2a). The designers of life have based their creation on the similar principles, knowledge and observation that we have about humans; which is the typical information used in SB. Hence, the objective of the project is to understand how the virtual worm lives and to determine any new knowledge that was out of the scope of the virtual worm designers. Two different strategies were applied to understand the life of the virtual worm: the Topological analysis and the Functional analysis. Both strategies have their limitations and requirements but the combination of both has allowed us to understand rather well the life of the virtual worm.

Comparing the results obtained in the Topological analysis and Functional analysis we can confirm that both methods come to similar conclusions when they treat the same problem. 75% of effector proteins determined by the Topological analysis have been confirmed as effector proteins in the Functional analysis (Figures 4c, d, e and f). Moreover, the seed proteins provided by the virtual worm designers as requirement for the Topological analysis were also confirmed by the Functional analysis (Figure 4b). So, it is possible to comply with the requirements of the Topological analysis using the list of seed proteins obtained from the Functional analysis. This fact is rather significant because it means that it is not necessary to have a broader knowledge about a species to reach some conclusions about its life. In this case, for instance, only the gene/protein map and some observations about its life were sufficient to describe the key elements of the virtual worm's life.

## 4.2. Combination of Functional and Topological Analysis: future perspective and possible applications

By means of the Functional analysis a mathematical model of the virtual worm was obtained. It is possible to obtain some answers from this mathematical algorithm that contains a summary of the virtual worms' life: how can I design a drug to treat a specific pathology? Can I understand the mechanism of action of a pathway? How can I foster the growth of this species? So, through the analysis of the mathematical algorithm it would be possible to design a specific drug for each specific individual or to predict any adverse event from the known drug targets of a drug. Is this strategy applicable to humans? Not yet, since some problems may occur in this case. Today, the known protein map for humans is composed by around 11.000 nodes or proteins, including interactome, metabolome and signaling pathways. The problem is to find the set of virtual chromosomes that explain all the complexity of humans and represent the solution of the mathematical model. The number of nodes and links in humans is so high that with the actual capacity for calculation it is not possible to find all the possible solutions and consequently it is not possible to carry out this exercise. A possible solution to these problems is to simplify the information by aggregating the proteins or increasing the abstraction of the nodes [15].

However, there are other important fields in health science where to apply this type of techniques, for instance, the investigation related to the simplest organisms: bacteria. Bacteria have relatively simple exome and the knowledge about them is ever more extensive. Therefore it is relatively easy to create a map of bacteria. If we apply the aforementioned techniques, we could design new antibiotics or identify new mechanisms of infection of bacteria.

There are several other examples of toy models to perform analysis of virtual life in order to design the best strategy to understand real life. Taking into account the amount of data generated by genome analysis and other biological disciplines, models for simulating natural processes have emerged. There is an enormous interest in this area and some disciplines have been trying to model real life, from purely computational models to biological models. Moreover, the level of representation goes from molecular levels (intracellular reactions [23] or cells) to animals and ecosystems where each artificial organism is designed as an autonomous agent and variables as deformable body or learning capacities are included in the model [24]. Even the ecosystems' behavior has been represented in a computational way to study the competition and adaptation to environment of two species [25]. From more biological perspective [26], the best modeling lies in an integrative approach. As example, if we want to model a cell, gene regulation, metabolism and environmental stimulation/response have to be included; even the simplest metabolic pathway is a part of a complex system with which it is connected. Therefore, it is necessary to find out how proteins work collectively as a living system.

## 5. ACKNOWLEDGEMENTS

This research leading to these results was co-funded by MICINN (EUI-2009-04018), partners of the ERASysBio+ initiative supported under the EU ERA-NET Plus scheme in FP7, and by European Union's Seventh Framework Programme (FP7/2007-2013) under grant agreements nº HEALTH-F3-2009-223101 (AntiPathoGN) and nº HEALTH-F2-2010-261460 (GUMS&JOINTS).

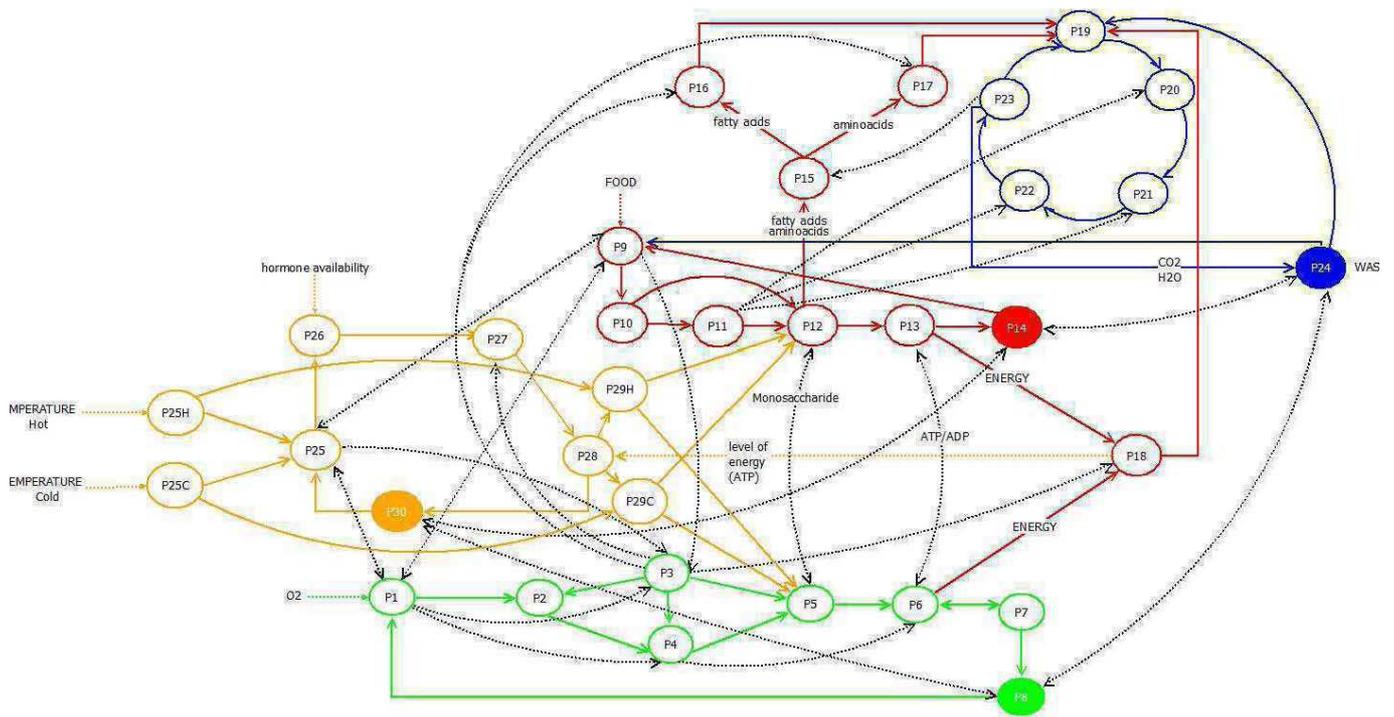

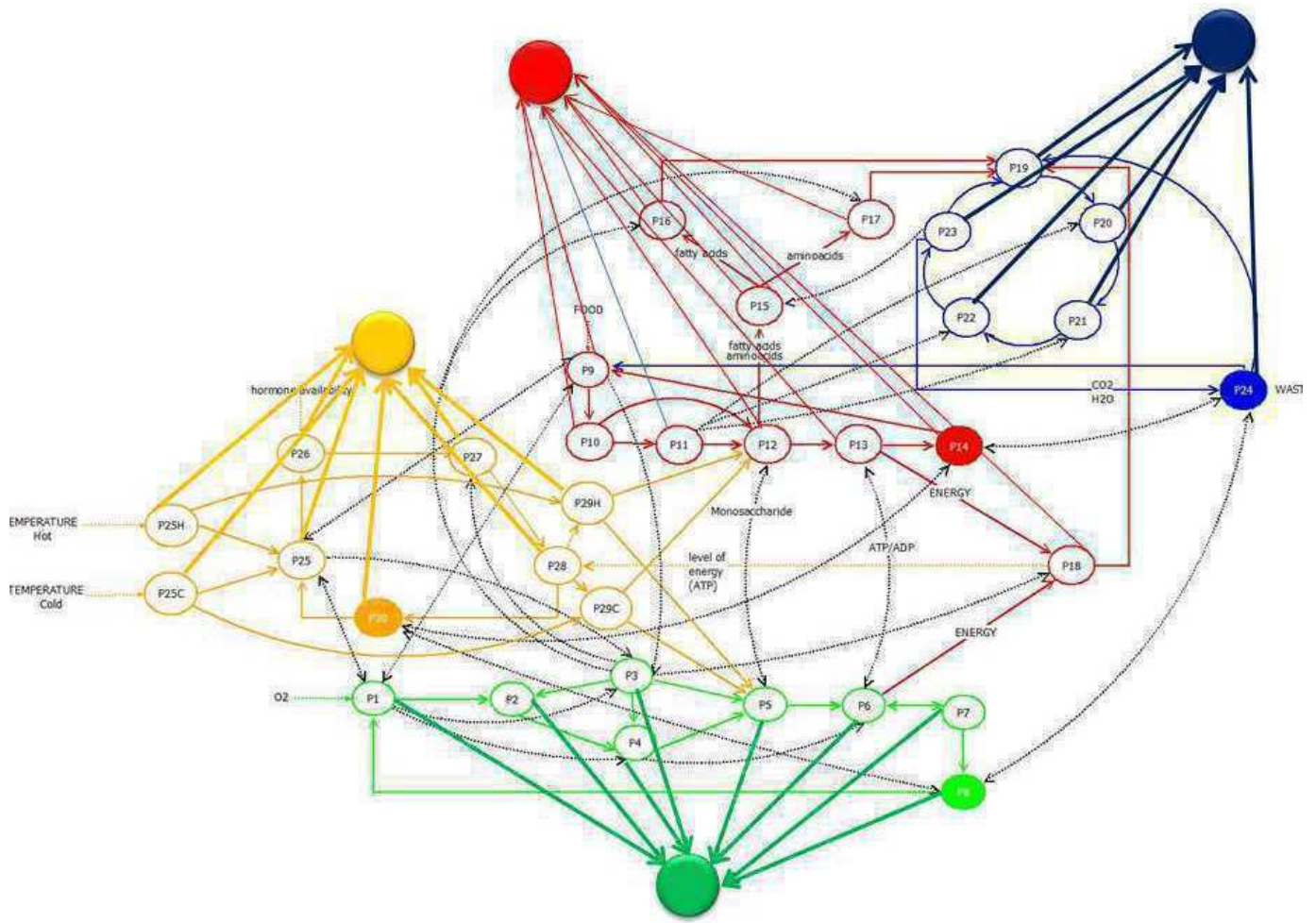

|  | Available information (INPUTS) | Hidden information | Results (OUTPUTS) | Validation |
|---|---|---|---|---|
| Topological Analysis | • Topology of the map<br>• Seed proteins* | Effector proteins** defined by the virtual worm designer. | Effector proteins** | Effector proteins** given by the virtual worm designer. |
| Functional Analysis | • Topology of the map<br>• Truth Table | • Seed proteins* defined by the virtual worm designer.<br>• Effector proteins** defined by the virtual worm designer. | • Effector proteins**<br>• Seed proteins* | Seed proteins* given by the virtual worm designer. |

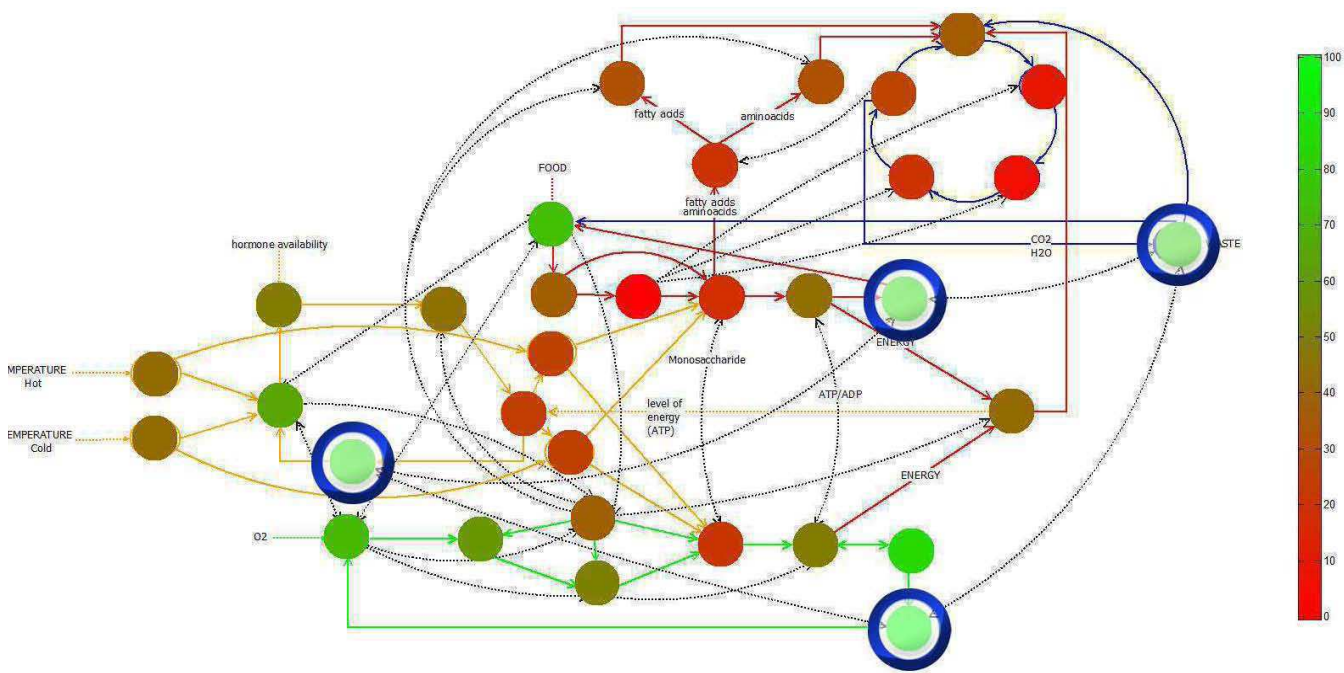

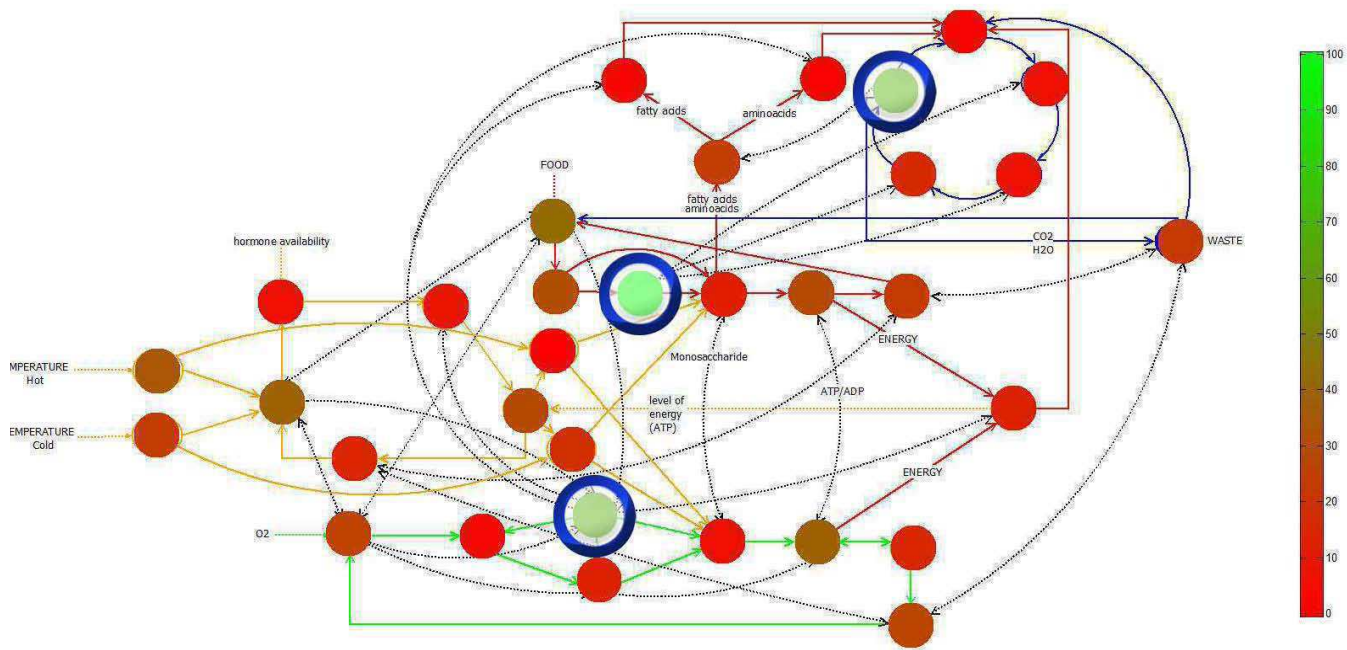
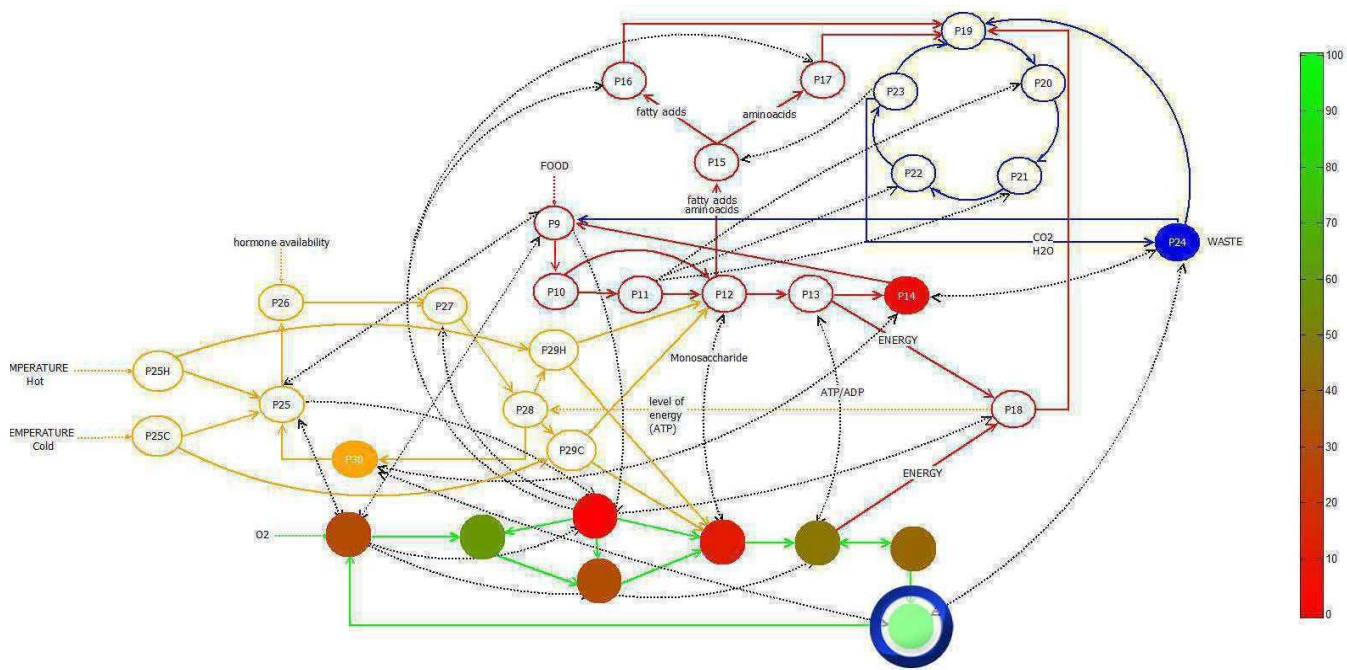

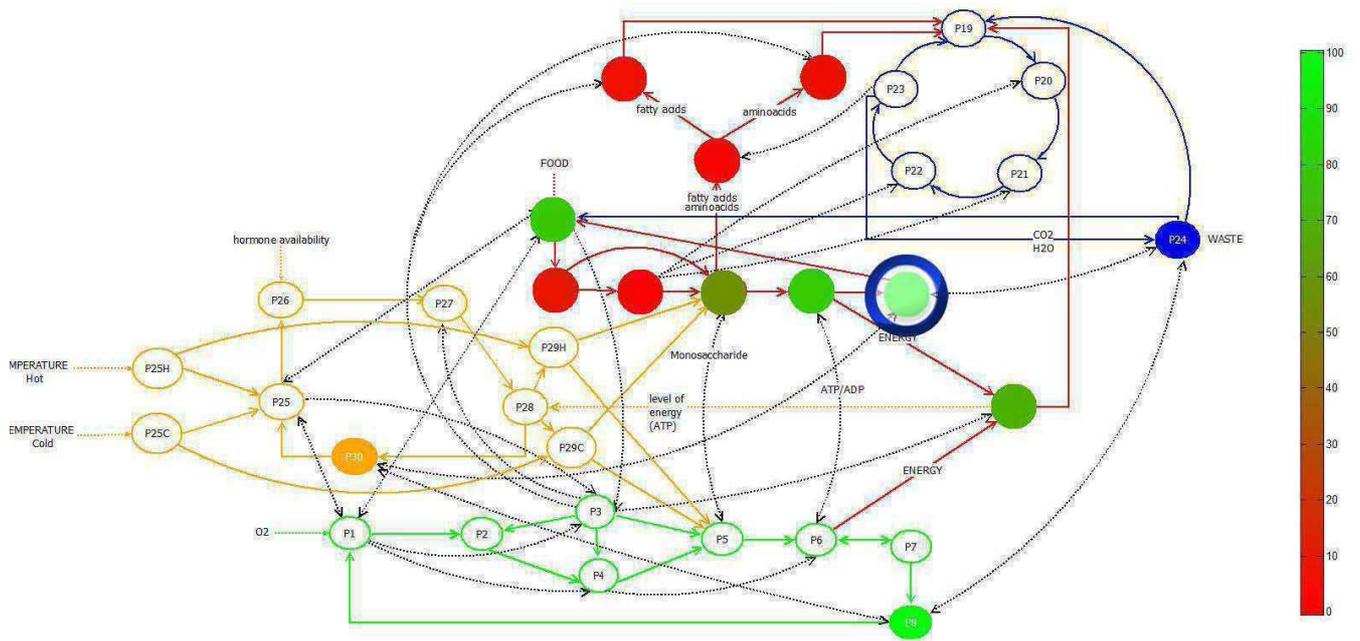
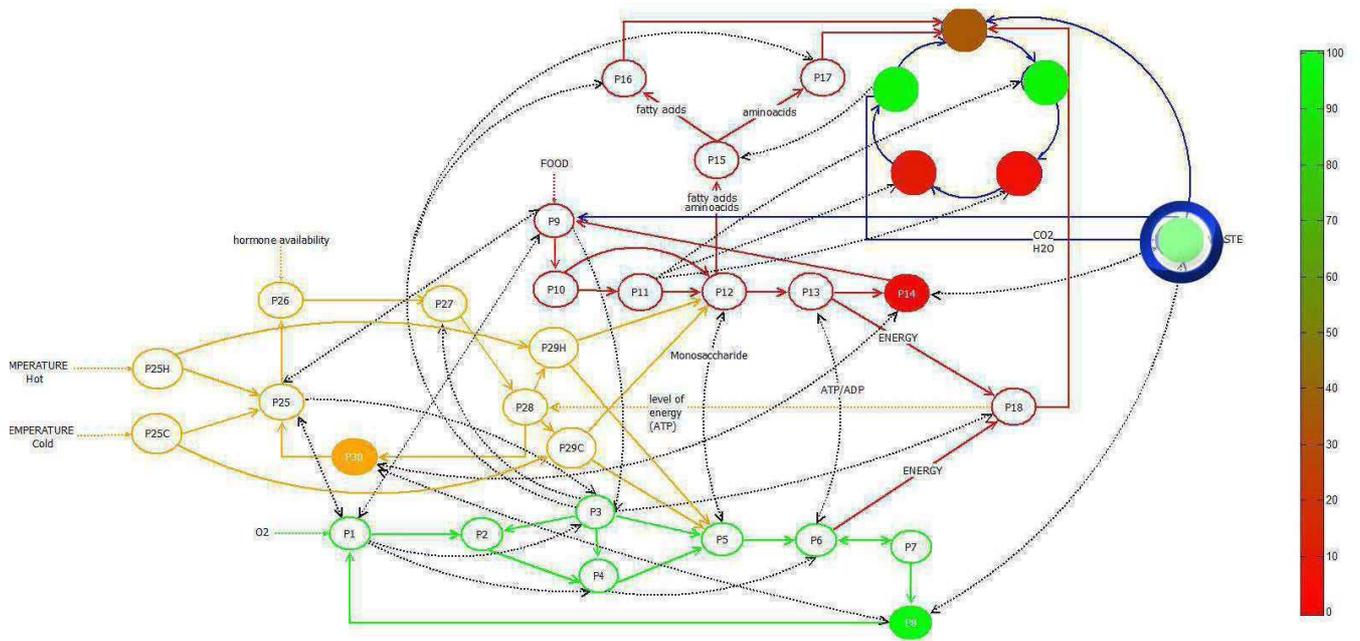

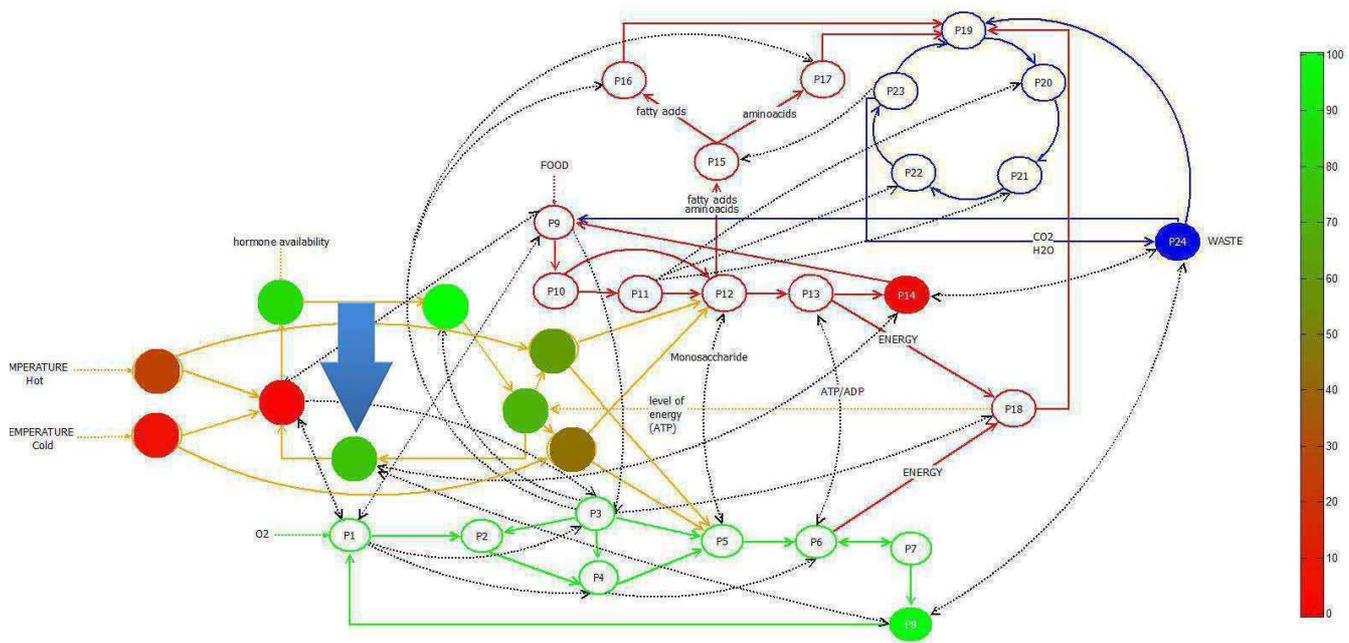